# Membrane fouling: microscopic insights into the effects of surface chemistry and roughness

Mao Wang[a], John Wang,[b] Jianwen Jiang[a*]

[a] Department of Chemical and Biomolecular Engineering, National University of Singapore, 117576, Singapore
[b] Department of Materials Science and Engineering, National University of Singapore, 117575, Singapore

**Abstract:** Fouling is a major obstacle and challenge in membrane-based separation processes. Caused by the sophisticated interactions between foulant and membrane surface, fouling strongly depends on membrane surface chemistry and morphology. Current studies in the field have been largely focused on polymer membranes. Herein, we report a molecular simulation study for fouling on alumina and graphene membrane surfaces during water treatment. For two foulants (sucralose and bisphenol A), the fouling on alumina surfaces is reduced with increasing surface roughness; however, the fouling on graphene surfaces is enhanced by roughness. It is unravelled that the foulant-surface interaction becomes weaker in the ridge region of a rough alumina surface, thus allowing foulant to leave the surface and reducing fouling. Such behavior is not observed on a rough graphene surface because of the strong foulant-graphene interaction. Moreover, with increasing roughness, the hydrogen bonds formed between water and alumina surfaces are found to increase in number as well as stability. By scaling the atomic charges of alumina, fouling behavior on alumina surfaces is shifted to the one on graphene surfaces. This simulation study reveals that surface chemistry and roughness play a crucial role in membrane fouling, and the microscopic insights are useful for the design of new membranes towards high-performance water treatment.

# 1. Introduction

In recent years, water scarcity has become a major concern worldwide due to ever-growing population, climate change and water pollution [1,2]. Clean water is indispensable for global health and economic development. It has become a grand challenge to provide clean water in a reliable and affordable manner. As an energy efficient and cost effective technology, membrane-based water treatment has a wide variety of applications in seawater or brackish water desalination, municipal or industrial wastewater purification. However, one of the key issues in membrane-based water treatment is membrane fouling. It adversely affects membrane performance by reducing water permeability, deteriorating water quality, lowering membrane lifetime, and increasing energy consumption [3,4].

Microscopically, fouling is caused by sophisticated interactions between membrane surface and foulant, which strongly depends on the chemistry and morphology of membrane surface. In this respect, surface modification is widely adopted to improve membrane anti-fouling capacity [5–7]. At present, most studies in this field have primarily investigated the effect of surface chemistry on fouling [8]. Nevertheless, surface morphology or roughness also has a significant impact. A number of studies have shown that surface roughness can enhance fouling. For example, Elimelech et al. found that colloidal particles were preferentially accumulated in the valleys of rough aromatic polyamide thin-film composite membranes, which resulted in "valley clogging" and severely reduced water flux [9,10]. Based a mathematical model, they explained that the valleys created by surface roughness produced a lower interaction well in which colloidal particles might preferentially deposit [11]. Using both experimental and simulation techniques, Jun et al. investigated the fouling of poly(methyl methacrylate) (PMMA) colloidal particles on polyvinylidene fluoride (PVDF) membrane surface with periodically spaced prism shape. A stagnant flow zone was found to form in the valley of patterned surface with more particles deposited [12]. On three commercial polyamide thin-film composite reverse osmosis (RO) membranes with different roughness and hydrophilicity, Yin et al. observed that the fouling increased with an increase in surface roughness [13].

Interestingly, the opposite effect of surface roughness on fouling (i.e., reduce fouling) was also reported. Gohari et al. fabricated flat sheet poly(ether sulfone) (PES) membranes with aligned patterns by incorporating hydrophilic nanoparticles for the ultrafiltration of bovine serum albumin (BSA). When feed solution flowing perpendicular to the patterns, they observed that the fouling could be mitigated; however, the fouled surface was washed more effectively for parallel feed solution [14]. Using a newly developed layered interfacial polymerization, Choi et al. produced a sharklet-patterned membrane, which exhibited remarkably lower biofouling compared to conventional membranes with irregular roughness and simple patterns [15]. The sharklet patterned membrane was revealed to possess optimal anti-biofouling effect if the unit and pattern spacings were both 2 μm, as attributed to a balance between intrinsic biofouling propensity and surface flow characteristics such as vortex and primary/secondary flows [16]. Through the Derjaguin-Landau-Verwey-Overbeek (DLVO) theory and density functional theory (DFT) calculations, Li et al. examined the effect of surface roughness on membrane fouling caused by alginate adhesion. A rough membrane was found to be less alginate adhesion and adhesive fouling than a smooth counterpart [17].

As discussed above, fundamental understanding in the effect of surface roughness on membrane fouling is elusive and most of the current studies are focused on polymer membranes. In the past decade, ceramic membranes have been increasingly applied for industrial/municipal wastewater and drinking water treatment [18]. Compared to polymer membranes, ceramic membranes particularly alumina-based exhibit stronger mechanical strength, higher water permeability, better chemical resistance, as well as longer lifetime. Recently, Lyu et al. developed a surface-patterned alumina ceramic membrane with gradient porous structure by using a 3D-printing technology, and the line-patterned membrane showed a notable increase in water flux and a significantly enhanced antifouling ability [19]. On the other hand, carbon membranes like graphene have also received attention due to their unique advantages such as low toxicity, flexibility in storage and handling, easy processing and modification [20]. Practically, the surface roughness of alumina and graphene membranes has been revealed to significantly affect water transport and separation performance [21–24]; however, the impact on fouling is less clear.

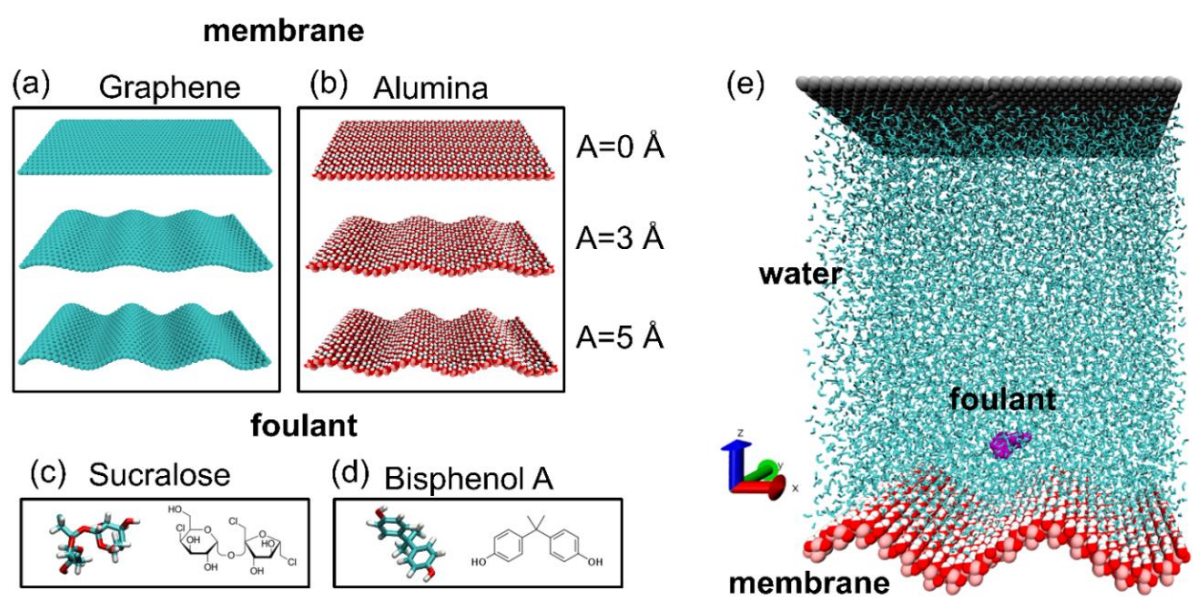

**Fig. 1.** (a-b) Graphene and alumina surfaces by varying roughness (amplitude A = 0, 3, 5 Å). (c-d) two model foulants: sucralose and bisphenol A. (e) A typical simulation system with a foulant molecule (purple) on an alumina surface (Al: pink; O: red; H: white). Water molecules are shown in cyan. A plate on the top is shown in grey.

Herein, a molecular simulation study is reported to investigate the effects of both surface chemistry and roughness on membrane fouling during water treatment. As illustrated in **Fig. 1a** and **1b**, we consider two types of membranes (hydrophilic alumina and hydrophobic graphene), thus representing different surface chemistry. For each type, the surface roughness is tuned by varying amplitude A (A = 0, 3 and 5 Å, respectively). Two organic molecules, sucralose (SUC) and bisphenol A (BPA), as illustrated in **Fig. 1c** and **1d** are used as foulants. SUC is a chlorinated carbohydrate, while BPA is a xenoestrogen [25]. Both foulants are commonly present in wastewater and have a similar size of about 1.2 nm. Following this introduction, the simulation models and methods are briefly described in Section 2. In Section 3, foulant mobility is first examined on the two types of membrane surfaces; then, the effects of surface chemistry and roughness on fouling are discussed in detail. Finally, the concluding remarks are summarized in Section 4.

2. Models and Methods

**Fig. 1e** illustrates a typical simulation system with a foulant molecule solvated in water about 12 Å from an alumina surface. The surface was fixed at the bottom of a simulation cell with a water compartment above. A plate (grey) was above the water compartment to separate it from the vacuum. To avoid the effect of the plate, the water compartment was sufficiently large of about 10 nm along the *z*-axis. The simulation cell was approximately 4 nm in both *x*- and *y*-axes. The alumina surface was considered to be fully hydroxylated with 15 hydroxyl group per nm$^2$ [26]. Along the *z*-axis, the alumina surface consisted of two layers of aluminum (Al) and oxygen (O) atoms underneath hydrogen (H) atoms [23,24].

For rough alumina and graphene surfaces, the roughness was introduced onto a flat surface by using a sinusoidal function [24]

$$z = A \sin \left(\frac{2\pi x}{40}\right) \tag{1}$$

were A is the wave amplitude, $x$ is the atomic position along the *x*-axis. For each type of surface, three different amplitudes (A = 0, 3 and 5 Å represented by A0, A3 and A5, respectively) were examined. The atoms in alumina and graphene were described by the nonbonded Lennard-Jones (LJ) and electrostatic potentials:

$$\sum 4\varepsilon_{ij} \left[ \left(\frac{\sigma_{ij}}{r_{ij}}\right)^{12} - \left(\frac{\sigma_{ij}}{r_{ij}}\right)^{6} \right] + \sum \frac{q_i q_j}{4\pi\varepsilon_0 r_{ij}} \tag{2}$$

where $\varepsilon_{ij}$ and $\sigma_{ij}$ are the potential well and collision diameter, $r_{ij}$ is the distance between atom *i* and *j*, $q_i$ is the charge of atom *i*, and $\varepsilon_0$ is the permittivity of vacuum. The LJ parameters and atomic charges of alumina were adopted from the CLAYFF force field [27], which is widely used for hydroxylated oxide materials such as alumina, aluminosilicates and silica. The LJ parameters of graphene were taken from the OPLS-AA force field [28]. Water was mimicked by the SPC/E model [29], because of its simplicity and capability to reproduce water diffusivity [30]. The LJ and bonded parameters of SUC and BPA were based on the AMBER force field [31], while their atomic charges were estimated from density functional theory calculations using GAUSSIAN 09 package [32].

Each system was first equilibrated by molecular dynamics (MD) simulation for 100 ps, with a pressure of 1 bar applied on the plate above the water compartment. Then, the MD simulation continued for 10 ns with the plate fixed. The simulations were performed at 300 K using velocity-rescaling thermostat and the time step was 2 fs. The electrostatic interactions were calculated by the Particle-Mesh Ewald method, while the LJ interactions were estimated using a cutoff of 1.2 nm. During the 10 ns MD simulation, the foulant molecule was free to move and its mean-squared displacement (MSD) was averaged from the second 5 ns trajectory:

$$\text{MSD}(t) = \left\langle \left| \Delta \mathbf{r}(t) \right|^2 \right\rangle \tag{3}$$

where $t$ is time and $\Delta \mathbf{r}(t)$ is the displacement of the foulant molecule. For each system, six independent runs were performed with different random seeds to generate initial velocities. In addition, the potentials of mean force (PMFs) for foulant on flat alumina graphene surfaces were calculated separately. More specifically, a force was applied to steer foulant molecule moving towards the flat surface. Umbrella sampling [33] was used involving 21 simulations with 0.05 nm interval along the $z$-axis. In each simulation, the position of foulant was restrained in the $z$-axis by a harmonic potential with a force constant of 8000 kJ mol$^{-1}$ nm$^{-2}$. The PMF was then generated from 3 ns simulation using the weighted histogram analysis method [34]. All the simulations in this study were conducted using GROMACS 5.1.4 [35].

## 3. Results and discussions

### 3.1. Foulant mobility

The MSD of foulant can be used to assess membrane fouling. When a foulant molecule is less mobile, it tends to deposit on membrane surface and leads to fouling [36,37]. **Fig. 2** shows the MSDs of SUC and BPA on alumina and graphene surfaces, respectively, by varying roughness. Each MSD curve is averaged over six independent runs, thus mitigating the effect of random diffusion. There are several interesting observations: (1) on the same surface (e.g., alumina A0), SUC has a smaller MSD than BPA, indicating more severe fouling. (2) On alumina surfaces, both foulants exhibit a similar trend of MSD with roughness; specifically, the MSD rises with increasing roughness (i.e., from A0, A3 to A5). This implies the foulant is more mobile and the fouling is reduced on a rougher alumina surface. (3) The opposite trend is observed on graphene surfaces for both foulants; with increasing roughness, the MSD drops and fouling is enhanced. These observations reveal that fouling is less dependent on the type of foulant, as both SUC and BPA possess similar fouling behavior on alumina or graphene surfaces; instead, the surface chemistry and roughness are dominant factors to determine fouling, which will be discussed in detail below.

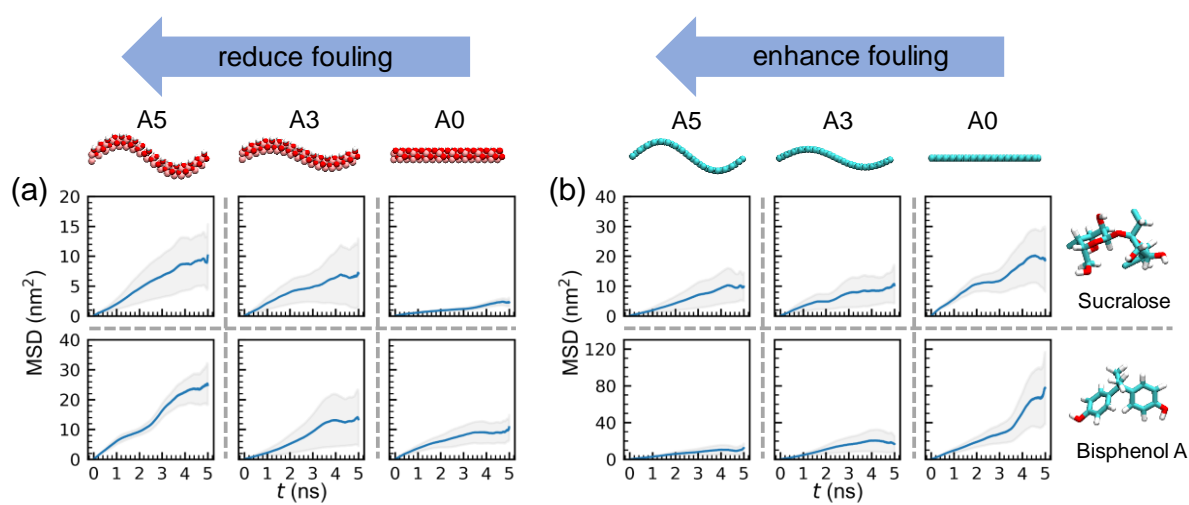

Fig. 2. MSDs of SUC and BPA on (a) alumina and (b) graphene and surfaces by varying roughness. Each MSD curve is averaged over six independent runs. The shaded region denotes statistical uncertainty. Note that the scales are different in (a) and (b).

## 3.2. Effect of surface chemistry

To quantitatively examine the effect of surface chemistry on fouling, we focus on the flat alumina and graphene surfaces. **Fig. 3** shows the PMFs of SUC and BPA as a function of distance for the alumina surface. While both foulants show a similar trend of PMF, SUC has a lower minimal PMF (-16.6 kJ mol$^{-1}$ at point 2) than BPA (-15.6 kJ mol$^{-1}$ at point 4). The lower PMF of SUC is consistent with the above smaller MSD of SUC on **Fig. 2a**, as a stronger interaction between foulant and surface causes less mobility. For each foulant, the second minimum is observed in the PMF profile (point 1 for SUC and point 3 for BPA). The four points (1, 2, 3 and 4) correspond to different preferred configurations when SUC and BPA are adsorbed on the alumina surface. At point 1, SUC is away from the surface, but comparatively closer to the surface at point 2; consequently, SUC interacts with the surface more strongly at point 2 than at point 1, as observed in the PMF profile. For BPA, one of its two ring is parallel to the surface at point 3 and the interaction is stronger than at point 4 without ring parallel to the surface.

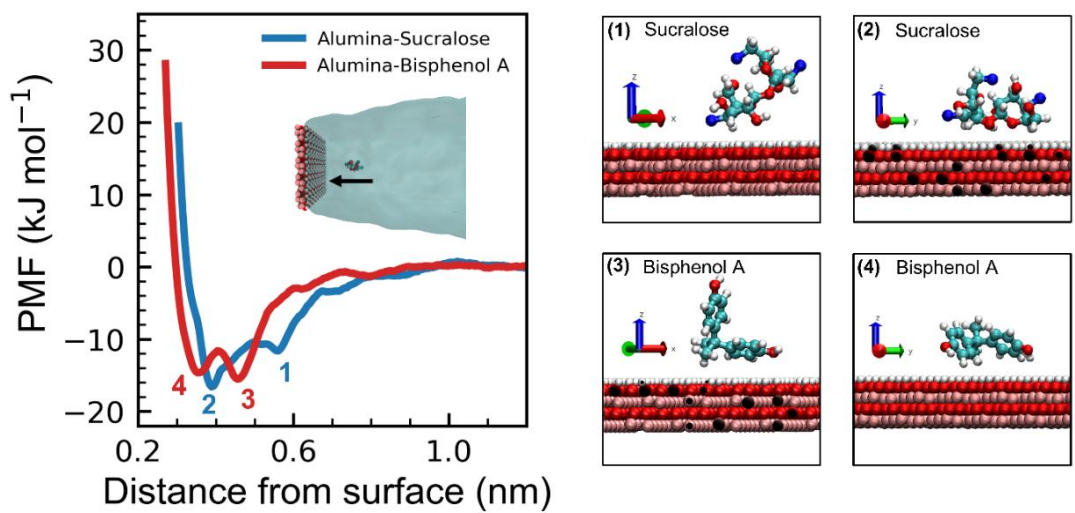

Fig. 3 (Left) PMFs of SUC and BPA on the flat alumina surface as a function of distance. The insert illustrates a foulant molecule moving towards the surface. (Right) Simulation snapshots at different distances (as labelled in the left figure) from the flat alumina surface.

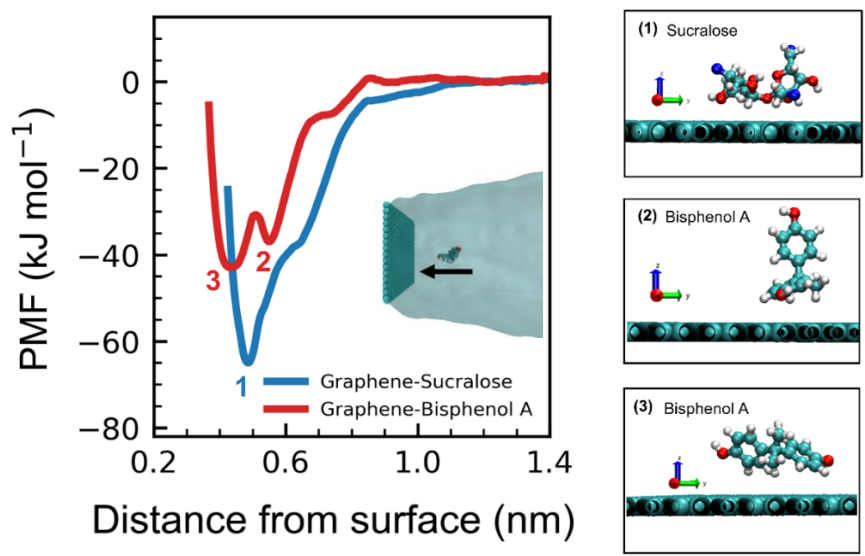

Fig. 4. (Left) PMFs of SUC and BPA on the flat graphene surface as a function of distance. The insert illustrates a foulant molecule moving towards the surface. (Right) Simulation snapshots at different distances (as labelled in the left figure) from the flat graphene surface.

On the flat graphene surface, there is distinct difference in the PMFs between SUC and BPA. As shown in **Fig. 4**, the minimal PMF is about -65 kJ mol$^{-1}$ for SUC at point 1, while it is -42.9 kJ mol$^{-1}$ for BPA at point 3. Meanwhile, the second minimum is seen at point 2 for BPA. The three points (1, 2 and 3) are attributed to different configurations on the graphene surface. From **Fig. 3** and **Fig. 4**, apparently, the PMFs of both foulants on the graphene surface are much deeper than on the alumina surface. The reason is that the hydrophobic graphene surface interacts more strongly with the two organic foulants compared to the hydrophilic alumina counterpart. Interestingly, as shown in **Fig. 2**, the MSDs of the two foulants appear to be larger on the graphene surface than on the alumina surface. With a closer look at the respective MSDs in the $z$ and $xy$ directions, as shown in **Fig. S1**, we can find that the foulant has a negligible MSD$_z$ but a significant MSD$_{xy}$. This is attributed to the atomic smooth surface of graphene, thus the foulant can move quite freely parallel to the surface. However, the movement perpendicular to the surface is largely prohibited due to strong interaction. Such a phenomenon was also observed in a simulation study for polymer chains adsorbed on the graphene surface [38].

## 3.3. Effect of surface roughness

We use a harmonic potential to steer a foulant molecule moving along different surfaces by varying roughness (see **Fig. 5a**). In this way, the foulant is used as a probe to quantify the foulant-surface interaction in the *x*-axis. **Fig. 5b** and **Fig. S2b** show the interaction energies of SUC and BPA with graphene surfaces, respectively. On the flat surface, the interaction remains similar despite certain fluctuation mainly due to the configuration change of foulant. Upon increasing surface roughness (from A0 to A3 to A5), a ridge-valley pattern is seen in the interaction profile, with a maximum in the valley and a minimum in the ridge. The rougher the surface, the stronger is the interaction in the valley. This trend is observed for both foulants. Thus, the surface roughness has a significant impact on foulant-surface interaction.

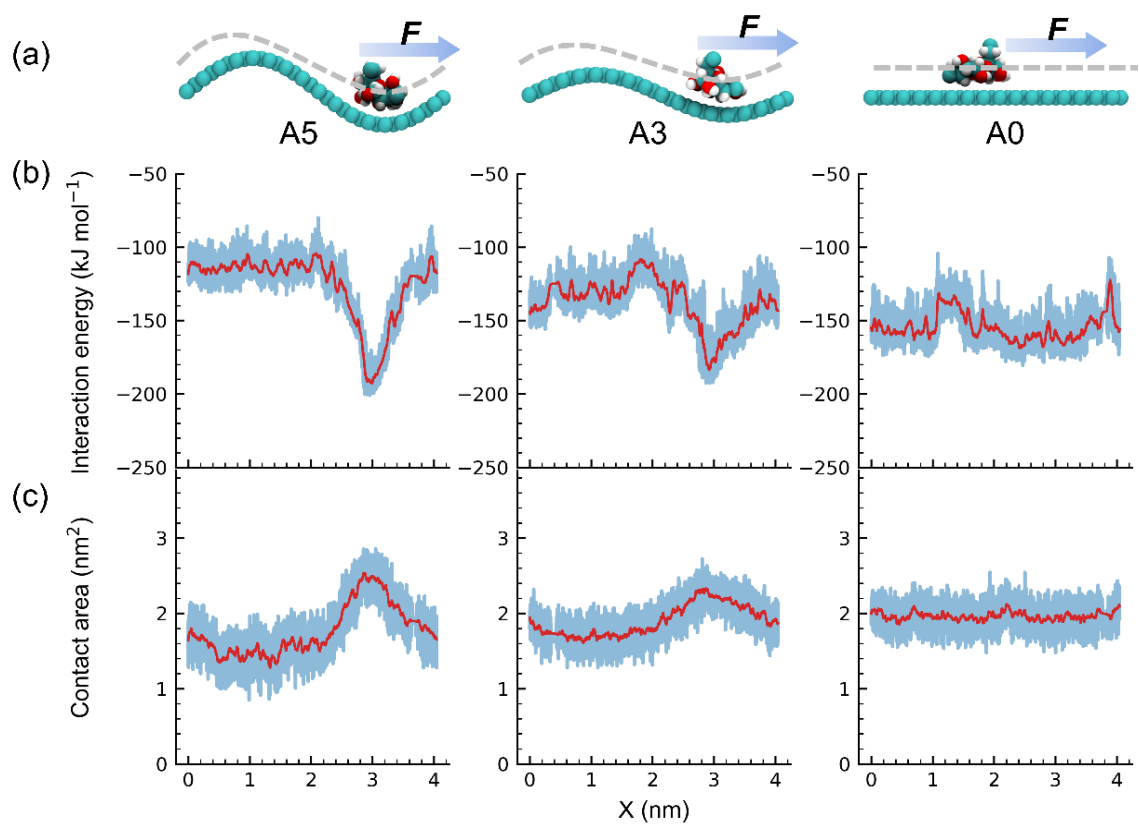

Fig. 5. (a) SUC moving along graphene surfaces under a harmonic force in the *x*-axis. (b, c) Interaction energies and contact areas between SUC and graphene surfaces.

The different interactions in the valley and ridge regions are closely related to the variation in foulant-surface contact areas. **Fig. 5c** and **Fig. S2c** show the contact areas between the two foulants and graphene surfaces. On the flat surface, the contact area is nearly a constant, as also seen in the interaction. On the rough surface, the contract area increases in the valley and accordingly, the interaction becomes stronger. This leads to "valley clogging" effect as experimentally observed [10]. By contrast, the contact area in the ridge decreases, which causes weaker foulant-surface interaction. In principle, the ridge is the place where foulant is most likely to depart from the surface. Nevertheless, the foulant interacts strongly with graphene surface; the interaction energies are about -110 kJ mol$^{-1}$ for SUC and -75 kJ mol$^{-1}$ for BPA even in the ridge of graphene surface A5, thus preventing foulant from leaving the surface. **Fig. S3** illustrates the snapshots of SUC in the valley and ridge of graphene surface A5. The foulant is observed to be closely adsorbed on the surface. Meanwhile, the adsorbed foulant can move parallelly along the graphene surface due to the atomic smooth surface, eventually the foulant can relocate from the ridge to the valley because the latter is more energetically favourable. As a consequence, fouling on graphene surfaces is enhanced by roughness, as evidenced by the MSDs in **Fig. 2**.

For alumina surfaces, the interaction energies of BPA and SUC are plotted in **Fig. 6b** and **Fig. S4b**, respectively. On the flat alumina surface, the interaction profile appears to be relatively constant. With increasing roughness, the interaction becomes stronger in the valley, similar to the case on rough graphene surface. However, the interaction in the ridge exhibits substantial difference between the two types of surfaces when roughness is sufficiently large. On the graphene surface A5, there is always strong interaction between foulant and surface in the ridge (see **Fig. 5b** and **Fig. S2b**) and the foulant is always adsorbed on the surface. On the alumina surface A5, however, the foulant may leave the surface in the ridge region as illustrated in **Fig. S5**, and hence the interaction may decrease to 0 (see **Fig. 6b** and **Fig. S4b**). **Fig. S6** shows the simulation snapshots of BPA on the alumina surface A5 without imposing a harmonic potential. The foulant is initially adsorbed on the surface at 1.4 ns, a bit away from the surface at 2.6 ns, and eventually leave the surface at 8.9 ns. **Fig. 6c** shows the contact areas between BPA and alumina surfaces. As also observed on graphene surfaces, the contact area increases in the valley and decreases in the ridge. On alumina surface A5, the contact area may become 0, which implies the foulant leaves the surface and the foulant-surface interaction decreases to 0 accordingly. From these results, apparently, fouling on alumina surfaces is reduced upon increasing roughness. This prediction is accord with a recent experimental study, in which a 3D-printed line-patterned alumina membrane was observed to be less fouling [19].

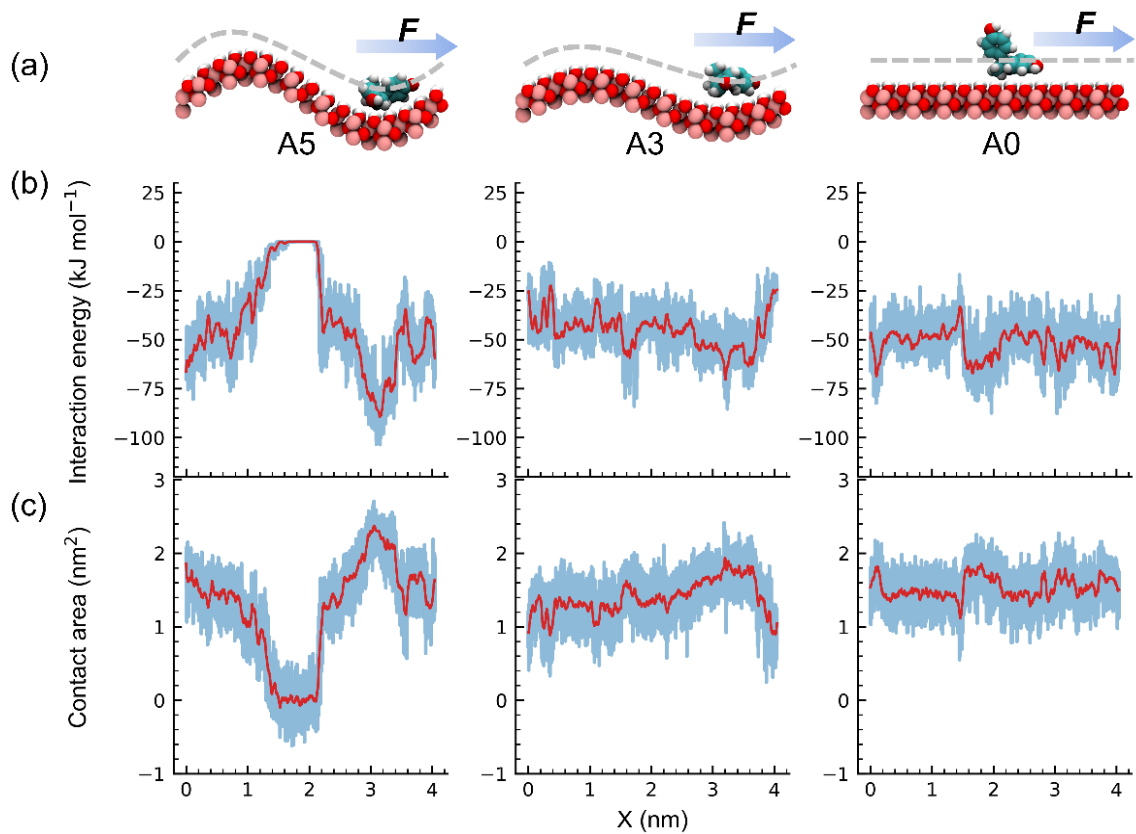

Fig. 6. (a) BPA moving along alumina surfaces under a harmonic force in the *x*-axis. (b, c) Interaction energies and contact areas between BPA and alumina surfaces.

To further quantify the effect of roughness on hydrophilic alumina surfaces, we analyse the hydrogen bonds between water and alumina surfaces as illustrated in **Fig. 7a**. **Fig. 7b** shows the numbers of hydrogen bonds on different alumina surfaces. With increasing roughness and hence increasing surface area, more hydrogen bonds are formed between water and the surface. The dynamics of hydrogen bonds can be evaluated by autocorrelation function:

$$c(t) = \frac{\langle h(t_0)h(t_0+t)\rangle}{\langle h(t_0)h(t_0)\rangle} \tag{4}$$

where $h(t) = 1$ if a hydrogen bond remains at time $t$ and $h(t) = 0$ if otherwise. The ensemble average $\langle \cdots \rangle$ is on all the hydrogen bonds formed. The $c(t)$ quantifies the lifetime of hydrogen bonds. From **Fig. 7c**, it is obvious that the $c(t)$ decays slower on a rougher alumina surface, which implies the hydrogen bonds are more stable and have a longer lifetime with increasing roughness. Thus, roughness increases not only the number of hydrogen bonds, but also the stability. Overall, the surface hydrophilicity of alumina surface is enhanced upon increasing roughness, which is reflected by the interaction energies between water and alumina surfaces in **Fig. 7d**.

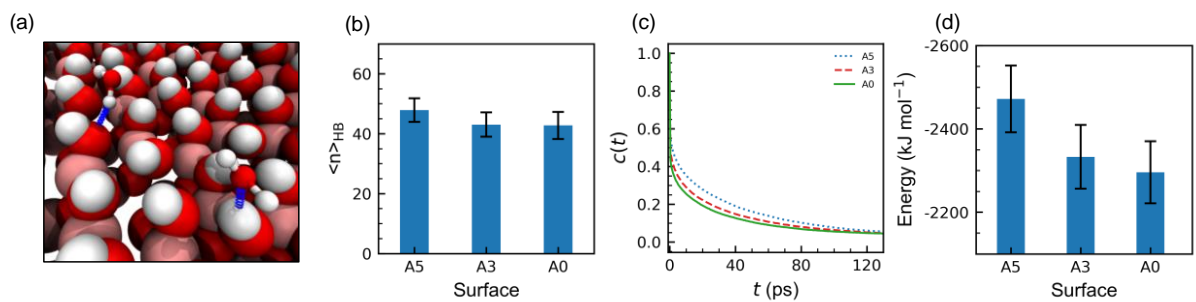

Fig. 7. (a) Hydrogen bonds (indicated by the blue lines) formed between water and alumina surfaces. Hydrogen: white, oxygen: red, aluminium: pink. (b-c) Numbers and autocorrelation functions of hydrogen bonds on alumina surfaces. (d) Interaction energies between water and alumina surfaces.

It is instructive to explore how the fouling behavior on hydrophilic alumina surfaces changes to that on hydrophobic graphene counterparts. To achieve this, we manipulate the hydrophilicity of alumina surfaces by scaling the atomic charges of alumina. For example, if a scaling factor is 0.5, the atomic charges are only 0.5 times of the actual alumina. Such a simple scaling method was used in the literature to change the hydrophilicity of MgO surface [39]. **Fig. S7** shows the interaction energies between water and alumina surfaces with different scaling factors. By decreasing the scaling factor, the water-surface interaction becomes weaker and apparently the surface is less hydrophilic. **Fig. 8** shows the MSDs of BPA on alumina surface with different scaling factors. On the actual alumina surfaces (i.e., scaling factor = 1), as shown in **Fig. 2a**, the MSD rises upon increasing roughness and hence fouling is reduced. When the scaling factor = 0.7, the MSD is nearly equal on A5 and A0 surfaces. When the scaling factor = 0.1, at which the surface is highly hydrophobic, the trend of MSD resembles that on graphene surfaces in **Fig. 2b**; i.e., the MSD drops on a rougher surface and fouling is enhanced. Clearly, these results reveal that the fouling behavior can be tailored by changing the hydrophilicity or hydrophobicity of surfaces.

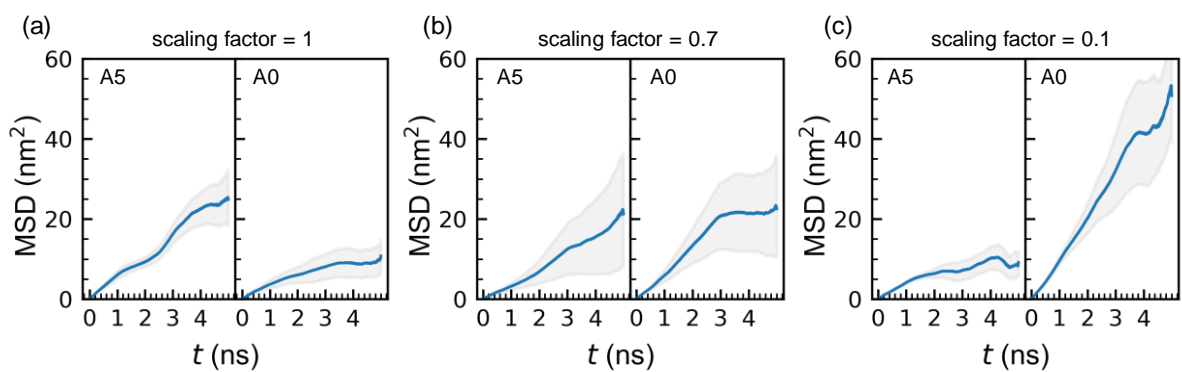

Fig. 8. MSDs of BPA on alumina surfaces with different scaling factors.

## 4. Conclusions

We have conducted molecular simulations to investigate the effects of surface chemistry and roughness on membrane fouling. For two foulants (sucralose and bisphenol A), the mobility on alumina surfaces is found to rise with increasing roughness and hence fouling is reduced by roughness. The opposite trend is observed on graphene surfaces, i.e., the mobility of foulant drops and fouling is enhanced with increasing roughness. These interesting effects of roughness on fouling are attributed to the different foulant-surface interactions and contact areas between alumina and graphene surfaces. It is also found that upon increasing roughness, more hydrogen bonds with longer lifetime are formed between water and alumina surface.

Furthermore, the fouling behavior can be readily tuned by varying the hydrophilicity/hydrophobicity of surfaces. The microscopic insights into the crucial effects of surface chemistry and roughness on fouling from this bottom-up study are useful for the development of anti-fouling membranes for water treatment and other important separation processes.

**Notes**

The authors declare no competing financial interest.

**Acknowledgements**

We gratefully acknowledge the Singapore National Research Foundation (NRF-CRP17-2017-01), the Singapore Ministry of Education and the National University of Singapore (R-279-000-598-114, R-279-000-574-114 and C-261-000-207-532/C-261-000-777-532) for financial support.

**References**

[1] M. Elimelech, W.A. Phillip, The Future of Seawater Desalination: Energy, Technology, and the Environment, Science. 333 (2011) 712–717. https://doi.org/10.1126/science.1200488.

[2] M.A. Shannon, P.W. Bohn, M. Elimelech, J.G. Georgiadis, B.J. Mariñas, A.M. Mayes, Science and technology for water purification in the coming decades, Nature. 452 (2008) 301–310. https://doi.org/10.1038/nature06599.

[3] R. Zhang, Y. Liu, M. He, Y. Su, X. Zhao, M. Elimelech, Z. Jiang, Antifouling membranes for sustainable water purification: strategies and mechanisms, Chemical Society Reviews. 45 (2016) 5888–5924. https://doi.org/10.1039/C5CS00579E.

[4] A. Al-Amoudi, R.W. Lovitt, Fouling strategies and the cleaning system of NF membranes and factors affecting cleaning efficiency, Journal of Membrane Science. 303 (2007) 4–28. https://doi.org/10.1016/j.memsci.2007.06.002.

[5] X. Zhao, R. Zhang, Y. Liu, M. He, Y. Su, C. Gao, Z. Jiang, Antifouling membrane surface construction: Chemistry plays a critical role, Journal of Membrane Science. 551 (2018) 145–171. https://doi.org/10.1016/j.memsci.2018.01.039.

[6] M. Asadollahi, D. Bastani, S.A. Musavi, Enhancement of surface properties and performance of reverse osmosis membranes after surface modification: A review, Desalination. 420 (2017) 330–383. https://doi.org/10.1016/j.desal.2017.05.027.

[7] Q. Gu, T.C.A. Ng, L. Zhang, Z. Lyu, Z. Zhang, H.Y. Ng, J. Wang, Interfacial diffusion assisted chemical deposition (ID-CD) for confined surface modification of alumina microfiltration membranes toward high-flux and anti-fouling, Separation and Purification Technology. 235 (2020) 116177. https://doi.org/10.1016/j.seppur.2019.116177.

[8] D.J. Miller, D.R. Dreyer, C.W. Bielawski, D.R. Paul, B.D. Freeman, Surface Modification of Water Purification Membranes, Angewandte Chemie International Edition. 56 (2017) 4662–4711. https://doi.org/10.1002/anie.201601509.

[9] M. Elimelech, Xiaohua Zhu, A.E. Childress, Seungkwan Hong, Role of membrane surface morphology in colloidal fouling of cellulose acetate and composite aromatic polyamide reverse osmosis membranes, Journal of Membrane Science. 127 (1997) 101–109. https://doi.org/10.1016/S0376-7388(96)00351-1.

[10] E.M. Vrijenhoek, S. Hong, M. Elimelech, Influence of membrane surface properties on initial rate of colloidal fouling of reverse osmosis and nanofiltration membranes, Journal of Membrane Science. 188 (2001) 115–128. https://doi.org/10.1016/S0376-7388(01)00376-3.

[11] E.M.V. Hoek, S. Bhattacharjee, M. Elimelech, Effect of Membrane Surface Roughness on Colloid−Membrane DLVO Interactions, Langmuir. 19 (2003) 4836–4847. https://doi.org/10.1021/la027083c.

[12] S.Y. Jung, Y.-J. Won, J.H. Jang, J.H. Yoo, K.H. Ahn, C.-H. Lee, Particle deposition on the patterned membrane surface: Simulation and experiments, Desalination. 370 (2015) 17–24. https://doi.org/10.1016/j.desal.2015.05.014.

[13] Z. Yin, C. Yang, C. Long, A. Li, Influence of surface properties of RO membrane on membrane fouling for treating textile secondary effluent, Environ Sci Pollut Res. 24 (2017) 16253–16262. https://doi.org/10.1007/s11356-017-9252-6.

[14] R. Jamshidi Gohari, W.J. Lau, T. Matsuura, A.F. Ismail, Effect of surface pattern formation on membrane fouling and its control in phase inversion process, Journal of Membrane Science. 446 (2013) 326–331. https://doi.org/10.1016/j.memsci.2013.06.056.

[15] W. Choi, C. Lee, D. Lee, Y.J. Won, G.W. Lee, M.G. Shin, B. Chun, T.-S. Kim, H.-D. Park, H.W. Jung, J.S. Lee, J.-H. Lee, Sharkskin-mimetic desalination membranes with ultralow biofouling, J. Mater. Chem. A. 6 (2018) 23034–23045. https://doi.org/10.1039/C8TA06125D.

[16] W. Choi, C. Lee, C.H. Yoo, M.G. Shin, G.W. Lee, T.-S. Kim, H.W. Jung, J.S. Lee, J.-H. Lee, Structural tailoring of sharkskin-mimetic patterned reverse osmosis membranes for optimizing biofouling resistance, Journal of Membrane Science. 595 (2020) 117602. https://doi.org/10.1016/j.memsci.2019.117602.

[17] R. Li, Y. Lou, Y. Xu, G. Ma, B.-Q. Liao, L. Shen, H. Lin, Effects of surface morphology on alginate adhesion: Molecular insights into membrane fouling based on XDLVO and DFT analysis, Chemosphere. 233 (2019) 373–380. https://doi.org/10.1016/j.chemosphere.2019.05.262.

[18] C. Li, W. Sun, Z. Lu, X. Ao, S. Li, Ceramic nanocomposite membranes and membrane fouling: A review, Water Research. 175 (2020) 115674. https://doi.org/10.1016/j.watres.2020.115674.

[19] Z. Lyu, T.C.A. Ng, T. Tran-Duc, G.J.H. Lim, Q. Gu, L. Zhang, Z. Zhang, J. Ding, N. Phan-Thien, J. Wang, H.Y. Ng, 3D-printed surface-patterned ceramic membrane with enhanced performance in crossflow filtration, Journal of Membrane Science. 606 (2020) 118138. https://doi.org/10.1016/j.memsci.2020.118138.

[20] A. Anand, B. Unnikrishnan, J.-Y. Mao, H.-J. Lin, C.-C. Huang, Graphene-based nanofiltration membranes for improving salt rejection, water flux and antifouling–A review, Desalination. 429 (2018) 119–133. https://doi.org/10.1016/j.desal.2017.12.012.

[21] Y. Kang, R. Qiu, M. Jian, P. Wang, Y. Xia, B. Motevalli, W. Zhao, Z. Tian, J.Z. Liu, H. Wang, H. Liu, X. Zhang, The Role of Nanowrinkles in Mass Transport across

Graphene-Based Membranes, Advanced Functional Materials. 30 (2020) 2003159. https://doi.org/10.1002/adfm.202003159.

[22] Y. Wei, Y. Zhang, X. Gao, Y. Yuan, B. Su, C. Gao, Declining flux and narrowing nanochannels under wrinkles of compacted graphene oxide nanofiltration membranes, Carbon. 108 (2016) 568–575. https://doi.org/10.1016/j.carbon.2016.07.056.

[23] A. Nalaparaju, J. Wang, J. Jiang, Enhancing water permeation through alumina membranes by changing from cylindrical to conical nanopores, Nanoscale. 11 (2019) 9869–9878. https://doi.org/10.1039/C8NR09602C.

[24] A. Nalaparaju, J. Wang, J. Jiang, Water Permeation through Conical Nanopores: Complex Interplay between Surface Roughness and Chemistry, Adv. Theory Simul. (2020) 2000025. https://doi.org/10.1002/adts.202000025.

[25] S.A. Vogel, The Politics of Plastics: The Making and Unmaking of Bisphenol A "Safety," Am J Public Health. 99 (2009) S559–S566. https://doi.org/10.2105/AJPH.2008.159228.

[26] K.C. Hass, W.F. Schneider, A. Curioni, W. Andreoni, The Chemistry of Water on Alumina Surfaces: Reaction Dynamics from First Principles, Science. 282 (1998) 265–268. https://doi.org/10.1126/science.282.5387.265.

[27] R.T. Cygan, J.-J. Liang, A.G. Kalinichev, Molecular Models of Hydroxide, Oxyhydroxide, and Clay Phases and the Development of a General Force Field, J. Phys. Chem. B. 108 (2004) 1255–1266. https://doi.org/10.1021/jp0363287.

[28] W.L. Jorgensen, D.S. Maxwell, J. Tirado-Rives, Development and Testing of the OPLS All-Atom Force Field on Conformational Energetics and Properties of Organic Liquids, J. Am. Chem. Soc. 118 (1996) 11225–11236. https://doi.org/10.1021/ja9621760.

[29] H.J.C. Berendsen, J.R. Grigera, T.P. Straatsma, The missing term in effective pair potentials, J. Phys. Chem. 91 (1987) 6269–6271. https://doi.org/10.1021/j100308a038.

[30] Z. Hu, J. Jiang, Assessment of biomolecular force fields for molecular dynamics simulations in a protein crystal, Journal of Computational Chemistry. 31 (2010) 371–380. https://doi.org/10.1002/jcc.21330.

[31] W.D. Cornell, P. Cieplak, C.I. Bayly, I.R. Gould, K.M. Merz, D.M. Ferguson, D.C. Spellmeyer, T. Fox, J.W. Caldwell, P.A. Kollman, A second generation force field for

the simulation of proteins, nucleic acids, and organic molecules, Journal of the American Chemical Society. 117 (1995) 5179–5197.

[32] Frisch, M. J.; Trucks, G. W.; Schlegel, H. B.; Scuseria, G. E.; Robb, M. A.; Cheeseman, J. R.; Zakrzewski, V. G.; Montgomery, J. A.; Stratmann, R. E.; Burant, J. C.; Dapprich, S.; Millam, J. M.; Daniels, A. D.; Kudin, K. N.; Strain, M. C.; Farkas, O.; Tomasi, J.; Barone, V.; Cossi, M.; Cammi, R.; Mennucci, B.; Pomelli, C.; Adamo, C.; Clifford, S.; Ochterski, J.; Petersson, G. A.; Ayala, P. Y.; Cui, Q.; Morokuma, K.; Malick, D. K.; Rabuck, A. D.; Raghavachari, K.; Foresman, J. B.; Cioslowski, J.; Ortiz, J. V.; Stefanov, B. B.; Liu, G.; Liashenko, A.; Piskorz, P.; Komaromi, I.; Gomperts, R.; Martin, R. L.; Fox, D. J.; Keith, T.; Al-Laham, M. A.; Peng, C. Y.; Nanayakkara, A.; Gonzalez, C.; Challacombe, M.; Gill, P. M. W.; Johnson, B. G.; Chen, W.; Wong, M. W.; Andres, J. L.; Head-Gordon, M.; Replogle, E. S.; Pople, J. A. Gaussian 09. Revision D.01 ed.; Gaussian.

[33] G.M. Torrie, J.P. Valleau, Nonphysical sampling distributions in Monte Carlo free-energy estimation: Umbrella sampling, Journal of Computational Physics. 23 (1977) 187–199. https://doi.org/10.1016/0021-9991(77)90121-8.

[34] J.S. Hub, B.L. de Groot, D. van der Spoel, g_wham—A Free Weighted Histogram Analysis Implementation Including Robust Error and Autocorrelation Estimates, J. Chem. Theory Comput. 6 (2010) 3713–3720. https://doi.org/10.1021/ct100494z.

[35] M.J. Abraham, T. Murtola, R. Schulz, S. Páll, J.C. Smith, B. Hess, E. Lindahl, GROMACS: High performance molecular simulations through multi-level parallelism from laptops to supercomputers, SoftwareX. 1–2 (2015) 19–25. https://doi.org/10.1016/j.softx.2015.06.001.

[36] M.B. Tanis-Kanbur, S. Velioğlu, H.J. Tanudjaja, X. Hu, J.W. Chew, Understanding membrane fouling by oil-in-water emulsion via experiments and molecular dynamics simulations, Journal of Membrane Science. 566 (2018) 140–150. https://doi.org/10.1016/j.memsci.2018.08.067.

[37] Y. Ma, S. Velioğlu, M.B. Tanis-Kanbur, R. Wang, J.W. Chew, Mechanistic understanding of the adsorption of natural organic matter by heated aluminum oxide particles (HAOPs) via molecular dynamics simulation, Journal of Membrane Science. 598 (2020) 117651. https://doi.org/10.1016/j.memsci.2019.117651.

[38] L. Xu, X. Yang, Molecular dynamics simulation of adsorption of pyrene–polyethylene glycol onto graphene, Journal of Colloid and Interface Science. 418 (2014) 66–73. https://doi.org/10.1016/j.jcis.2013.12.005.

[39] T.A. Ho, D.V. Papavassiliou, L.L. Lee, A. Striolo, Liquid water can slip on a hydrophilic surface, PNAS. 108 (2011) 16170–16175. https://doi.org/10.1073/pnas.1105189108.

# Supplementary information

**Membrane fouling: microscopic insights into the effects of surface chemistry and roughness**

Mao Wang[a], John Wang,[b] Jianwen Jiang[a*]

[a] Department of Chemical and Biomolecular Engineering, National University of Singapore, 117576, Singapore
[b] Department of Materials Science and Engineering, National University of Singapore, 117575, Singapore

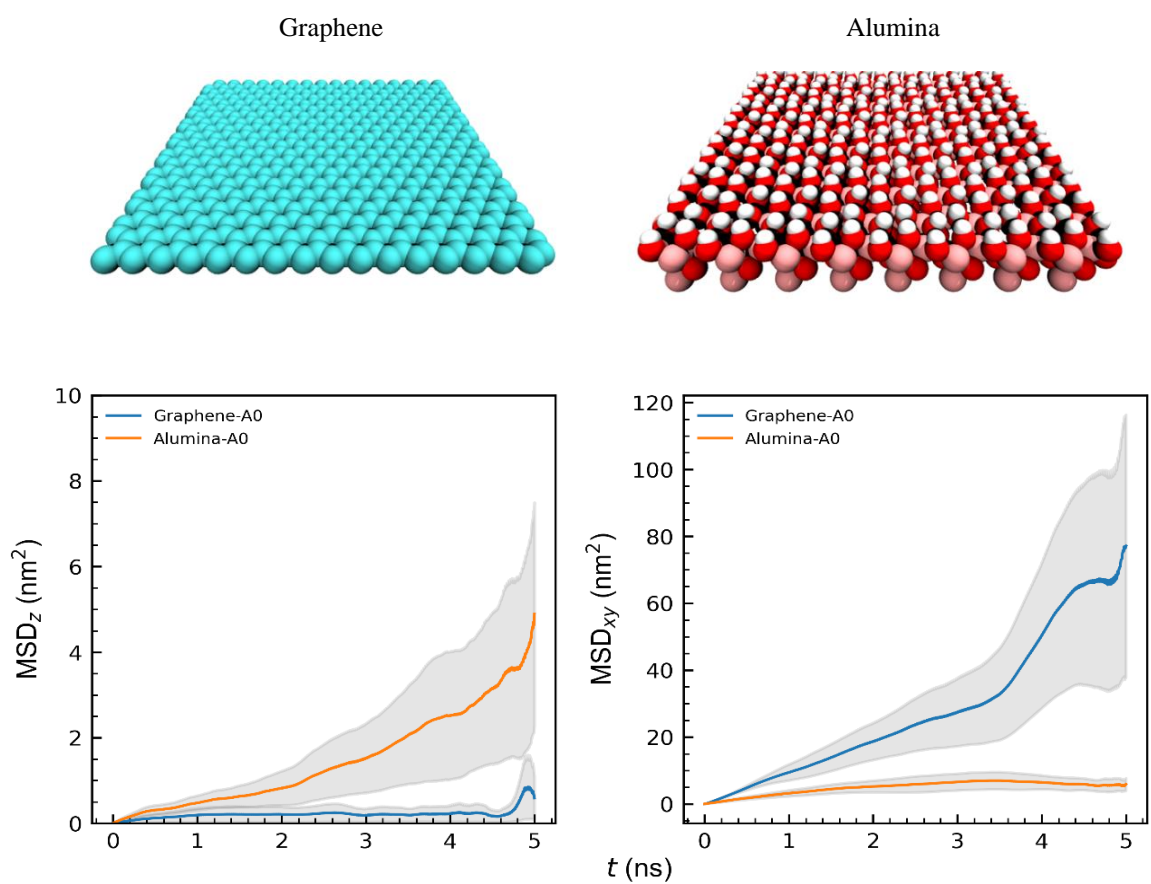

Fig. S1. MSDs of BPA in the *z* and *xy* directions on the flat graphene and alumina surfaces. The shaded region denotes statistical uncertainty.

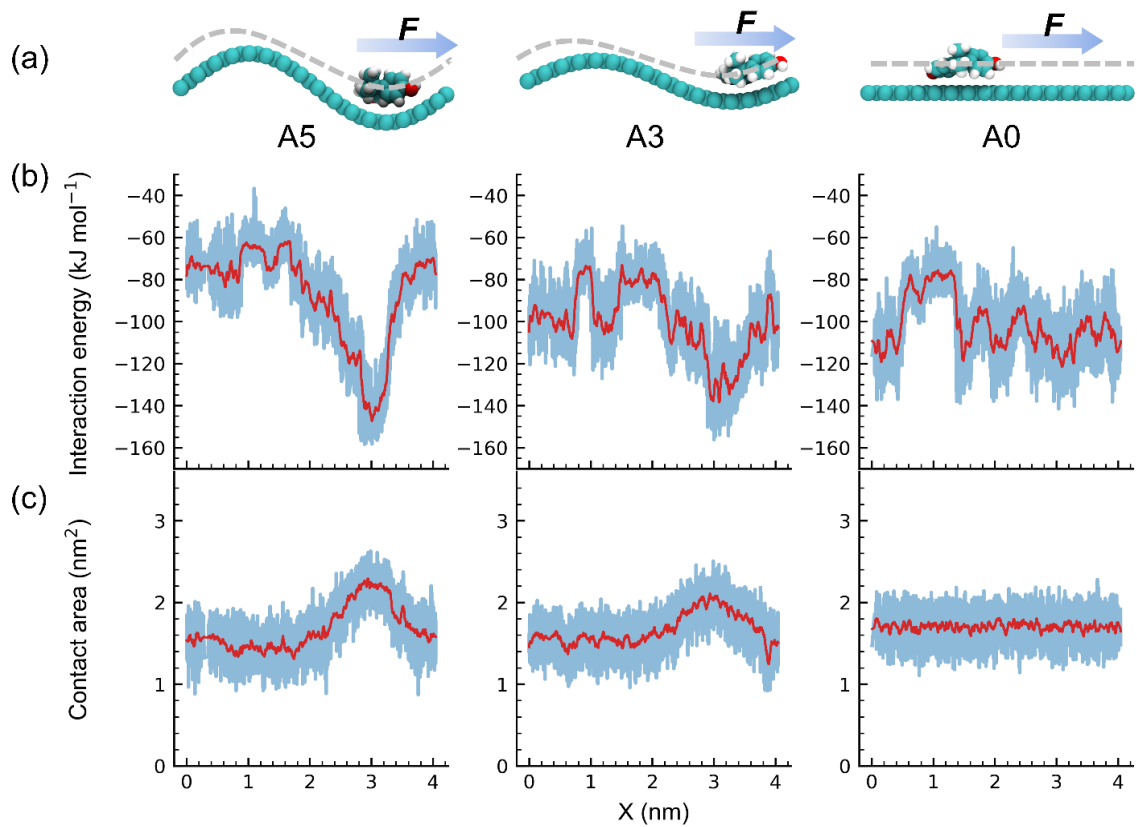

Fig. S2. (a) BPA moving along graphene surfaces under a harmonic force in the *x*-axis.

(b, c) Interaction energies and contact areas between BPA and graphene surfaces.

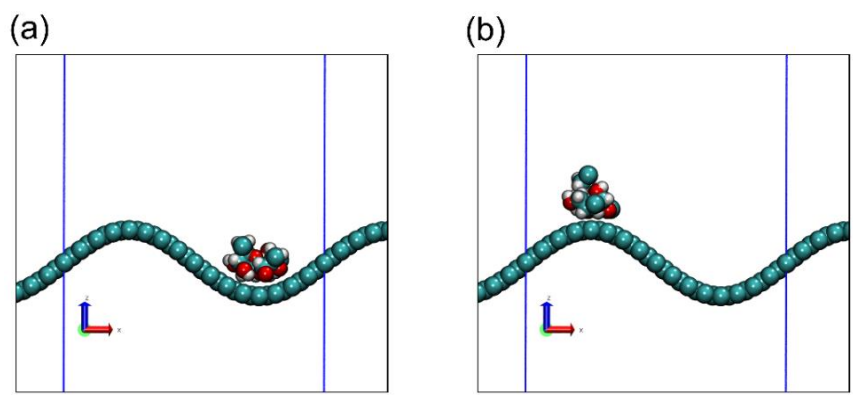

Fig. S3. Snapshots of SUC in the (a) valley and (b) ridge of graphene surface A5.

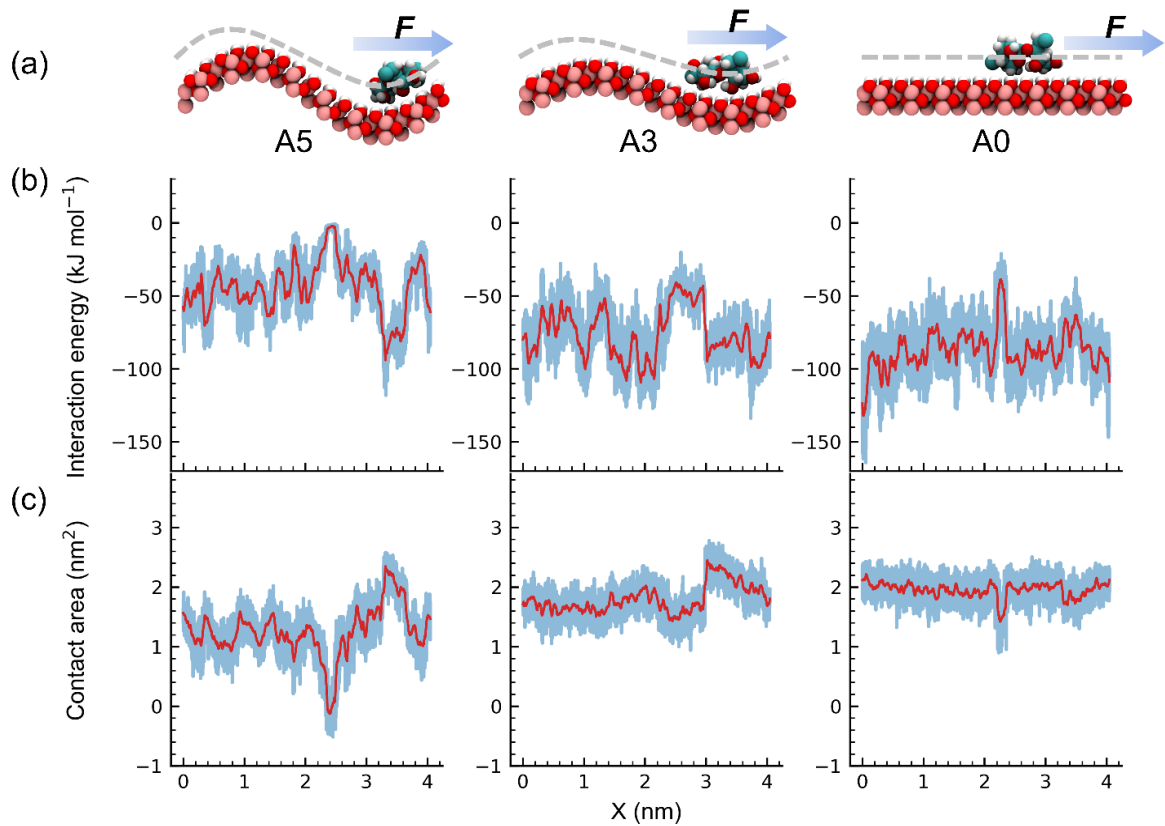

Fig. S4. (a) SUC moving along alumina surfaces under a harmonic force in the $x$-axis. (b, c) Interaction energies and contact areas between SUC and alumina surfaces.

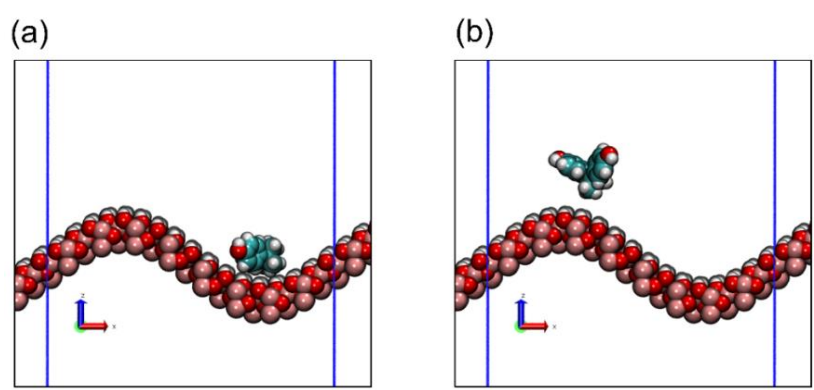

Fig. S5. Snapshots of BPA in the (a) valley and (b) ridge of alumina surface A5.

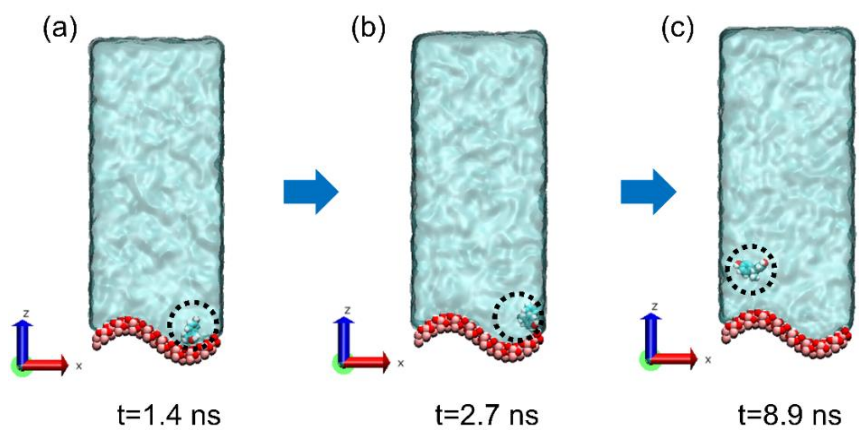

Fig. S6. Simulation snapshots of BPA on alumina surface A5 at different times.

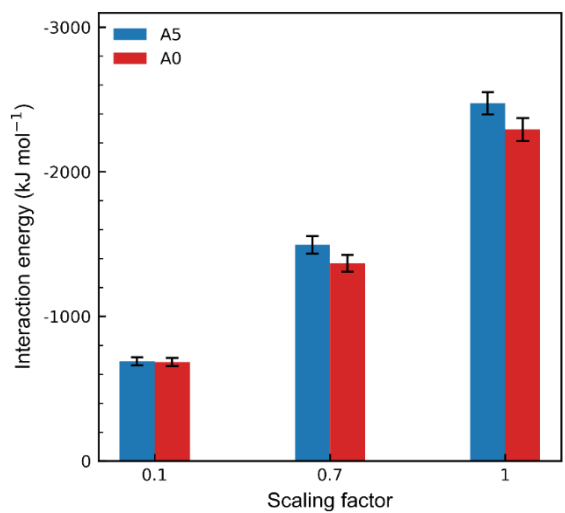

Fig. S7. Interaction energies between water and alumina surfaces with different scaling factors.